\documentclass[twocolumn,prb,aps,nobibnotes]{revtex4}
\usepackage[dvips]{graphicx}
\usepackage{amssymb,amsfonts,amsmath}
\usepackage{bm}
\usepackage{dcolumn}
\begin{document}

\title{\centering\Large\bf 
  Glassy protein dynamics and gigantic solvent reorganization
  energy of plastocyanin}
\author{David N.\ LeBard}
\author{Dmitry V.\ Matyushov}
\affiliation{Center for Biological Physics, 
         Arizona State University, PO Box 871604, Tempe, AZ 85287-1604}

\begin{abstract}
  We report the results of Molecular Dynamics simulations of electron
  transfer activation parameters of plastocyanin metalloprotein
  involved as electron carrier in natural photosynthesis. We have
  discovered that slow, non-ergodic conformational fluctuations of the
  protein, coupled to hydrating water, result in a very broad
  distribution of donor-acceptor energy gaps far exceeding that
  observed for commonly studied inorganic and organic donor-acceptor
  complexes. The Stokes shift is not affected by these fluctuations
  and can be calculated from solvation models in terms of the response
  of the solvent dipolar polarization. The non-ergodic character of
  large-amplitude protein/water mobility breaks the strong link
  between the Stokes shift and reorganization energy characteristic of
  equilibrium (ergodic) theories of electron transfer.  This mechanism
  might be responsible for low activation barriers in natural electron
  transfer proteins characterized by low reaction free energy.
\end{abstract}
\maketitle

\section{Introduction}
\label{sec:0}
\begin{figure}[t]
  \centering
  \includegraphics*[width=7cm]{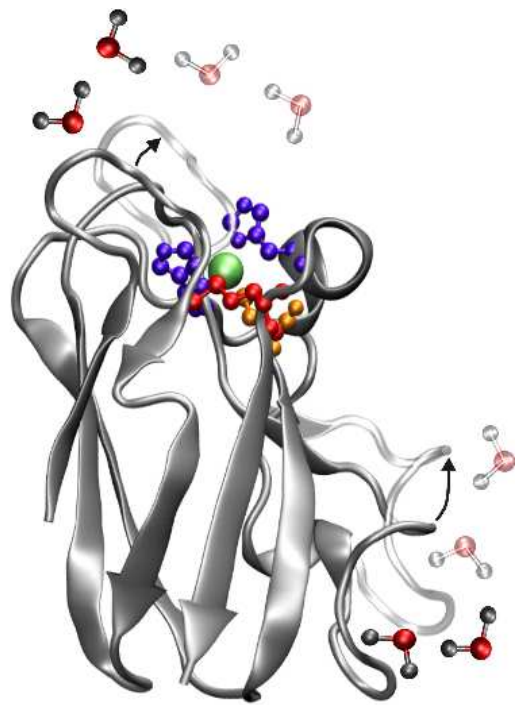}
  \caption{Structure of plastocyanin and the illustration of the
    protein large-scale conformational motions displacing hydrating
    water. The active site includes copper ion (green), 2 histidines
    (blue), methionine (red), and cysteine (orange) residues. The
    arrows and transparent parts of the protein illustrate motions of
    the main chain loops (not from actual MD simulations) displacing
    water molecules. }
  \label{fig:1}
\end{figure}

Redox proteins play diverse roles as electron carriers in biological
energy chains.\cite{Gray:05} Enzymatic activity often involves
transferring electrons to carry chemical reactions,\cite{Warshel:01}
while metalloproteins deposited in mitochondrial membranes and
photosynthetic units serve as redox sites with tuned redox potential
to allow one-directional electron flow in electron transfer
chains.\cite{Noy:06} Plastocyanin (PC) from spinach is a single
polypeptide chain of 99 residues forming a $\beta$-sandwich, with a single
copper ion coordinated by 2 sulfurs from cysteine and methionine and 2
nitrogens from histidine residues (Figure \ref{fig:1}). The presence of
the copper ion, which can change redox state, allows PC to function as
a mobile electron carrier in the photosynthetic apparatus of plants
and bacteria. It accepts an electron from ferrocytochrome \textit{f}
and diffusionally carries it to another docking location where the
electron is donated to the oxidized form of Photosystem
I.\cite{Ubbink:98}

This functionality is achieved through fast electron transfer
reactions at docking locations with low driving force $\simeq 20$ meV and
electron tunneling distance $>10$ \AA.\cite{Ubbink:98} The efficient
turnover of the photosynthetic apparatus demands fast rates at redox
sites, faster than typical biological catalytic rates of $10^2-10^4$
s$^{-1}$ (ref \onlinecite{Noy:06}). Given the small driving force,
this constraint limits the reorganization energy $\lambda$ to about 1
eV.\cite{Gray:00} The reorganization energy here is a sum of the
solvent, $\lambda_s$, and internal, $\lambda_i$, components, where $\lambda_s$ generally
incorporates the combined electrostatic effect of the protein and
water. In the rest of the paper, we will separate the atoms with
partial charges varying with the redox state as the redox site (Figure
\ref{fig:1}), considering the rest of the protein and water as the
thermal bath. Our main focus will, however, be on the interaction of
the redox site with water and that is how we define the solvent
reorganization energy $\lambda_s$ separating the interactions with the
protein atomic charges into the protein reorganization energy
$\lambda_{\mathrm{prot}}$ (see below for more precise definition).

Both $\lambda_i$ and $\lambda_s$ are large in synthetic redox systems with a
copper ion serving as the redox site because of large structural
changes upon electron transfer and a strong electrostatic interaction
of copper with polar solvents. Experimental
measurements\cite{Solomon:04} and quantum
calculations\cite{Cascella:06} of the internal reorganization energy
are still inconclusive placing it between 0.1 eV\cite{Cascella:06} and
0.6--0.7 eV (ref \onlinecite{Solomon:04} and references therein).  In
addition, recent numerical simulations of heme and copper proteins,
have uniformly placed their solvent reorganization energies in the
range of 0.5--1.0
eV.\cite{Warshel:01,Cascella:06,Muegge:97,Olsson:03,Blumberger:06,Sulpizi:07}
These calculations give results somewhat higher than what follows from
the experimental work on Ru-modified \textit{aeruginoza} azurin which
has shown that the activation barrier disappears at $\lambda\simeq 0.6-0.8$
eV.\cite{Skov:98} Even if the internal reorganization energy is as low
as 0.1 eV, the available data suggest that electronic transitions
involving copper proteins are significantly constrained
thermodynamically requiring a tight docking
configuration\cite{Hoffman:05} and strong electronic overlap within
the donor-acceptor pair which can therefore be modulated by protein's
conformations.\cite{Skourtis:05} One therefore wonders if there are
possibly some mechanisms at play, which are not included in standard
models of electron transfer,\cite{MarcusSutin} but which might allow a
greater tolerance in varying the parameters affecting the activation
barrier.  Our simulations reported here in fact show that the
combination of charged surface residues with the coupled protein/water
dynamics\cite{Tarek_PhysRevLett:02,Bizzarri:04,Chen:05} leads to a
lower activation barrier without requiring either a larger driving
force or a higher electronic overlap. PC is used here as a prototype
of what may apply to many other proteins involved in electron transfer
chains given the wide spread of the type of protein/water fluctuations
considered here among other proteins not necessarily involved in redox
activity.\cite{Fenimore:04}

\section{Energetics of electron-transfer activation}
\label{sec:1}
Electron transfer reactions are driven by thermal fluctuations of the
nuclear degrees of freedom interacting with the electronic states of
the donor and acceptor. Electrostatic interactions between the atomic
partial charges of the redox site with the partial charges or
multipoles of the thermal bath usually follow the rules of the linear
response approximation\cite{Blumberger:06,Kuharski:88,Simonson:02} embodied in
Marcus theory of electron transfer.\cite{MarcusSutin} The activation
barrier is calculated in this picture from the crossing of two
parabolic free energy surfaces $G_i(X)$ depending on the
donor-acceptor energy gap $X$. The use of equilibrium statistical
mechanics to calculate $G_i(X)$ results in several fundamental
equations between the (spectroscopically\cite{Marcus:89}) observable
parameters of the model.  The difference between equilibrium vertical
energy gaps ($\Delta X$, Stokes shift) is equal to twice the reorganization
energy $\lambda_p$ and is also related to the variance of the energy gap
$\sigma_s^2=\langle (\delta X)^2\rangle$ (spectral width):
\begin{equation}
  \label{eq:1}
  \Delta X = X_{01} - X_{02} = 2\lambda_p = \sigma_s^2 / k_{\mathrm{B}} T
\end{equation}
In eq \ref{eq:1}, $\lambda_p$ refers to the solvent reorganization energy
which, in traditional theories, is associated with the solvent
polarization field.\cite{MarcusSutin} This new notation is used here
to distinguish the traditional definition of the solvent
reorganization energy from our results for the solvent reorganization
energy $\lambda_s$ discussed below, which includes a new component not
identified in the previous studies. In addition to eq \ref{eq:1}, energy
conservation within Boltzmann statistics requires a linear
relation:\cite{Hwang:87}
\begin{equation}
  \label{eq:13}
   G_2(X)=G_1(X)+X  
\end{equation}
Many attempts,\cite{Kuharski:88} including those for redox
proteins,\cite{Cascella:06,Blumberger:06,Simonson:02} to test eqs
\ref{eq:1} and \ref{eq:13} have given positive results validating the
picture of two crossing parabolas. In contrast, our results here
report a breakdown of both relations by coupled protein-water
fluctuations affecting the statistics of the donor-acceptor energy
gap.

\begin{figure}
  \centering
  \includegraphics*[width=6cm]{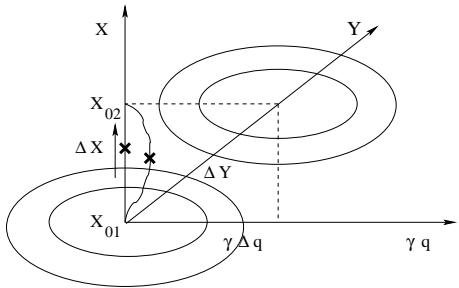}
  \caption{Electron transfer activation in two-coordinate space
    including solvent coordinate $X$ and classical conformational
    coordinate $q$ (multiplied by the factor $\alpha$ of electron-phonon
    interaction). The line connecting the two minima of the
    two-dimensional paraboloids corresponds to the reaction path $Y$
    for the fully thermalized fluctuations of both $X$ and $q$
    coordinates.  The total Stokes shift $\Delta Y$ is then the energetic
    distance between the minima.  Slow non-ergodic fluctuations of $q$
    shift the reaction path from the straight line connecting two
    equilibrium points $X_{0i}$ to the wiggled line.  The transition
    state then shifts from the cross point on the $X$-axis to a new
    point on the wiggled line.}
  \label{fig:2}
\end{figure}

Protein electron transfer adds the protein matrix as an additional
thermal bath characterized by a spectrum of vibrational modes
including fast molecular vibrations incorporated into the internal
reorganization energy $\lambda_i$ and slow conformational modes
affecting the electrostatic potential at the redox
site.\cite{Frauenfelder:91,Gehlen:94,McMahon:98,Liebl:99,Min:05} This
complication requires describing the reaction kinetics in terms of a
multidimensional reaction coordinate space. The physics of the
classical motions in the system is captured by a two-dimensional
paraboloid energy surface
\cite{Agmon:83,Sumi:86,Walker:92,Hoffman:96,Medvedev:06} as a function
of classical solvent, $X$, and effective vibrational, $q$, reaction
coordinates (Figure \ref{fig:2}). When both modes are fully
equilibrated on the reaction time-scale, the reaction path $Y=X+\gamma q$
is given as a linear combination of $X$ and $q$ with $\gamma$ representing
the electron-phonon coupling.  This reaction path dissects the
two-dimensional space along the line connecting the minima of two
paraboloids. The energetic separation between the minima defines the
full Stokes shift $\Delta Y$ related to the overall thermal dissipation of
the energy of electronic excitations by the thermal bath. Extending eq
\ref{eq:1} to equilibrium statistics in multidimensional coordinate
space, one can obtain $\Delta Y$ in terms of the reorganization energies:
\begin{equation}
  \label{eq:12}
  \Delta Y = 2(\lambda_p + \lambda_i)
\end{equation}
However, when one of the modes is slow, the reaction path deflects
from the line connecting the two minima and follows the fast reaction
coordinate.  The final state of the reaction then falls on the
$X$-axis and is denoted by $X_{02}$ in Figure \ref{fig:2}. This
picture, in which the solvent is a fast mode and the solute
conformational mobility is a slow coordinate, was first considered by
Agmon and Hopfield.\cite{Agmon:83} The problem of two-dimensional
dynamics was later formalized by Sumi and Marcus who focused, in
contrast, on the opposite case of fast intramolecular
vibrations.\cite{Sumi:86}

The fully equilibrated path along the coordinate $Y$ represents the
lowest potential barrier between the two equilibrium points.  If
conformational equilibrium is not achieved on the reaction timescale
$\tau_{\mathrm{ET}}= k_{\mathrm{ET}}^{-1}$ ($k_{\mathrm{ET}}$ is the
electron transfer rate), the reaction follows the path along $X$ with
the transition state marked by the cross on the $X$-axis (Figure
\ref{fig:2}). However, if the reaction path deviates from the straight
line due to stochastic conformational motions of the protein, it
potentially can pass through a lower transition state marked by the
cross on the wiggled line. The result of this is the breakdown of the
link between the Stokes shift along the coordinate $X$, given as $\Delta
X$, and the effective curvature of the free energy surface determined
by the variance of the energy gap fluctuations $\sigma_s^2 =\langle (\delta X)^2\rangle$ (eq
\ref{eq:1}).

The modulation of the donor-acceptor energy gap by protein motions can
be modeled by \textit{stochastic noise} in contrast to equilibrium
distribution resulting in eq \ref{eq:12}. This effect is accounted for
by adding average over conformational fluctuations (subscript ``q'')
to the Gaussian distribution along the solvent reaction coordinate
\begin{equation}
  \label{eq:2}
   e^{-G_i(X)/k_{\mathrm{B}}T} \propto 
        \left\langle \exp\left[- \frac{(X-X_{0i}(q))^2}{4k_{\mathrm{B}}T\lambda_p}\right]\right \rangle_q
\end{equation}
Here, the dependence on $q$ comes to the vertical energy gap
$X_{0i}(q)= \mathbf{P}_{\mathrm{eq},i}(q)*\Delta \mathbf{E}_0 $ (the
asterisk refers to both the scalar product and space
integration). This gap is formed by equilibrium solvent (nuclear)
polarization $\mathbf{P}_{\mathrm{eq},i}(q,\mathbf{r})$ in response to
all partial charges of the protein (1439 atomic charges for PC) and
the difference electric field $\Delta \mathbf{E}_0(\mathbf{r})$ created by
the difference charges $\Delta z_j$ of the redox site (Table S1, $j$ runs
over the atoms of the redox site).  On the contrary, the
polarization reorganization energy $\lambda_p$ is calculated as the
solvation free energy of $\Delta z_j$ charges only, and is affected by
positions of only a few atoms of the active site (4 in our
simulations, see Sec.\ \ref{sec:2}).  Therefore, one can expect that
it is the vertical gap that is predominantly modulated by protein
motions while $\lambda_p$ is mostly insensitive to such fluctuations (see
below).

Assuming Gaussian statistics of $\delta q$ and a linear expansion of
$X_{0i}(q)$ in $\delta q$ ($X_{0i}(q)\simeq X_{0i} + F \delta q$), one gets:
\begin{equation}
  \label{eq:3}
   G_i(X) = G_{0i} + \frac{(X - X_{0i})^2}{4(\lambda_p + \lambda_q(\tau_{\mathrm{obs}}))}
\end{equation}
The new reorganization energy $\lambda_q(\tau_{\mathrm{obs}})$ in principle
carries the dependence on the redox state ($i=1,2$), which requires
non-parabolic energy surfaces\cite{DMacc:07} and is not considered
here.  

The reorganization energy $\lambda_q(\tau_{\mathrm{obs}})$ carries the
dependence on the observation time $\tau_{\mathrm{obs}}$ in order to
stress on its non-ergodic character\cite{Hoffman:96,DMacc:07}
contrasting with equilibrium averages referring to $\tau_{\mathrm{obs}}\to
\infty$. The necessity to consider nonergodic activation parameters arises
from the wide spectrum of relaxation times typical of proteins.
$\lambda_q(\tau_{\mathrm{obs}})$ arises from the protein motions fast enough to
produce energy gap fluctuations on the time frame $\tau_{\mathrm{obs}}$
used to collect the averages. It can be obtained as the frequency
integral of the autocorrelation function $C_q(\omega)=\langle|\delta q_{\omega}|^2\rangle$ of $\delta
q_{\omega}$ with the low-frequency cutoff reflecting the final observation
time \cite{DMacc:07}
\begin{equation}
  \label{eq:8}
  \lambda_q((\tau_{\mathrm{obs}}) = 
             (F^2/k_{\mathrm{B}}T) \int_{\tau_{\mathrm{obs}}^{-1}}^{\infty} C_q(\omega) d\omega 
\end{equation}
It turns into equilibrium reorganization energy $\lambda_q=F^2/(2\kappa) $ ($\kappa$
is an effective force constant of harmonic conformational motions) in
the limit $\tau_{\textrm{obs}}\to\infty$ considered by statistical mechanics.

The conformationally-induced variance of the donor-acceptor energy gap
\begin{equation}
   \label{eq:7}
   \sigma_q^2(T_{\mathrm{kin}}) = 2k_{\mathrm{B}}T_{\mathrm{kin}}\lambda_q(T_{\mathrm{kin}}) 
\end{equation}
is in principle accessible experimentally from heterogeneous
electron-transfer kinetics measured on proteins cryogenically quenched
in their conformational substates.\cite{McMahon:98} Here, the
temperature of kinetic arrest $T_{\mathrm{kin}}$ is estimated by
requiring that the quenching rate $Q=dT/dt$ and the temperature
derivative of the conformational relaxation time $\tau_q$ produce unity
in their product: $Q \times (d\tau_q / dT) = 1$.\cite{Hodge:94} This approach,
however, eliminates the hydration dynamics facilitating conformational
changes (see below). One can therefore expect that such experiments
will inevitably underestimate $\sigma_q^2$ observed at high
temperatures.

The arguments presented in this section are not meant to give an
accurate theoretical description of the complex non-ergodic kinetics
of electron transfer influenced by protein/water dynamics. They are
more intended to set up a framework to understand the results of 
MD simulations which provide a more detailed picture of the nuclear
modes involved in the modulation of the donor-acceptor energy gap.

\section{Computational Methods}
\label{sec:2}
\subsection{MD Simulations}
\label{sec:2-1}
Amber 8.0\cite{amber8} was used for all MD simulations.  The initial
configuration of PC was created using X-ray crystal structure at 1.7 \AA{}
resolution (PDB: 1ag6\cite{1ag698}).  The system was heated in a NVT
ensemble for 30 ps from 0 K to the desired temperature followed by
volume expansion in a 1 ns NPT run. After density equilibration, NVT
production runs lasting from 15 ns (at 310 K) to 18 ns (at 285 K) were
made, of which 10 ns at the end of each trajectory were used to
calculate the averages.  The length of simulations was determined by
monitoring the convergence of the solvent reorganization energy $\lambda_s$,
which is the slowest-converging energetic parameter calculated
here. The timestep for all MD simulations was 2 fs, and SHAKE was
employed to constrain bonds to hydrogen atoms. Constant pressure and
temperature simulations employed Berendsen barostat and thermostat,
respectively.\cite{berend84} The long-range electrostatic interactions
were handled using a smooth particle mesh Ewald summation with a $9$ \AA{}
limit in the direct space sum.  The total charge for the protein was
$-9.0$ for the reduced (Red) state and $-8.0$ for the oxidized (Ox)
state. Each state was neutralized with the corresponding number of
sodium ions and TIP3P model was used for water.\cite{tip3p:83}

Three atomic charging schemes were utilized to parametrize PC's redox
site (Table~S1).  For the first parameter set, a chemically fake
charging scheme was employed that uses typical Amber force field
(FF03\cite{amberFF03}) for all standard amino acid residues, but
assigns an integer charge to the copper center in the reduced and
oxidized states (Q1).  Second, a more accurate charging scheme was
based upon experimental spin densities from Solomon's group for the
copper and copper ligands.\cite{Solomon:04,Ullmann:97} Finally, a
third charge distribution is completely parametrized at the DFT level
for the charges and force constants of the copper and ligand atoms and
consistent with the Amber force field (Q3).\cite{realq299} In
addition, Amber FF03 parametrization\cite{amberFF03} was applied to
all non-ligand residues (Q2).  There were various numbers of TIP3P
water molecules for each of the charge distributions: 5,874 (Q1),
5,886 (Q2), and 4,628 (Q3).

We ran separate simulations (ca.\ 5 ns) for each charging scheme to
find that the results are not strongly affected by the choice of
atomic charges.  This was also noticed in some other recent
simulations.\cite{Cascella:06,Blumberger:06} We have therefore
implemented charge scheme Q2 in all simulations reported here since it
presents a good balance between being simple and realistic.

Amber force field\cite{amberFF03} was also used for the ground state
tryptophan. Charges in the $^{1}L_{a}$ excited state were taken from
the literature.\cite{trpExcitedQ} This charge set was chosen because
it gives a good agreement with \textit{ab initio} calculations of the
indole ring.\cite{trpExcitedQ2} NVT simulations of tryptophan were
carried out for a total of 3 ns in a simulation box containing 420
water molecules after 1 ns density equilibration using NPT protocol
with a Berendsen barostat.\cite{berend84} The Stokes shift correlation
function\cite{Jimenez:94} simulated for tryptophan was in excellent
agreement with both the experimental data\cite{Pal:04} and previous
computer simulations.\cite{Nilsson:05} This model simulation was used
as a testing tool for our analysis of the Stokes shift dynamics of PC.

\subsection{Calculations of the solvation thermodynamics}
\label{sec:2-2}
Calculations of the solvent reorganization energy and the solvent part
of the reaction free energy gap were carried out by two methods: (i)
non-local response function theory (NRFT)\cite{DMjcp2:04,DMcp:06} and
(ii) dielectric continuum approximation implemented in the DelPhi
program suite.\cite{delphi02} Dielectric constant of TIP3P water
($\epsilon_s=97.5$\cite{DMcp:06}) was used for the solvent continuum and $\epsilon_s = 1$ for the
protein. This latter choice was driven by our desire to compare
continuum and microscopic calculations of solvation thermodynamics
since the latter does not assume any polarization of the protein. A
full account of the algorithm in application to the calculation of the
redox entropy of PC will be published elsewhere.\cite{DMjcp2:08} In
addition, since TIP3P water is non-polarizable $\epsilon_{\infty}=1.0$ was
used for the high-frequency dielectric constant in the reorganization
energy calculations.

In short, the NRFT calculation scheme employs the linear response
approximation to replace the solvation chemical potential with the
variance of the solute-solvent interaction
potential $V_{0s}$:\cite{DMjcp1:99,DMjpca:02}
\begin{equation}
  \label{eq:9}
  -\mu_{0s}= (2k_{\text{B}}T)^{-1} \langle \left(\delta V_{0s}\right)^2 \rangle_0
\end{equation}
The subscript ``0'' in the ensemble average $\langle \dots \rangle_0$ refers
to the fact that, in the linear response approximation, the spectrum
of electrostatic fluctuations of the solvent is not perturbed by the
electrostatic solute-solvent interactions. Therefore, the variance in
eq \ref{eq:9} is calculated for a fictitious system composed of water
solvent and the repulsive core of the solute with all solute charges
turned off. This approximation is known to work well for dense polar
solvents,\cite{DMjpca:02,DMjpcb1:03,DMcp:06} and the main problem of
the theory development is how to calculate the response function of
the polar solvent in the presence of the solute which expels the
dipolar polarization field from its volume.\cite{DMjcp1:04,DMjcp2:04}
This problem can be solved by applying the Gaussian solvation
model\cite{Chandler:93} resulting in the linear response function
(2-rank tensor) $\bm{\chi}[\bm{\chi}_s,\Omega_0]$ functionally depending on the
self-correlation function of the dipolar fluctuations of the solvent
$\bm{\chi}_s(\mathbf{k})$ and the shape of the solute occupying volume
$\Omega_0$. Once this problem is solved, the solvation chemical potential
is calculated as a 3D, inverted-space integral convoluting the
electric field of the solute $\mathbf{\tilde E}_0(\mathbf{k})$ with
the response function:
\begin{equation}
  \label{eq:10}
  -\mu_{0s}= \frac{1}{2} \mathbf{\tilde
    E}_0*\bm{\chi}[\bm{\chi}_s,\Omega_0]*\mathbf{\tilde E}_0^*
\end{equation}
Here, the asterisk refer to both the $\mathbf{k}$-integration and
tensor contraction and $\mathbf{\tilde E}_0^*$ is the complex conjugate of
$\mathbf{\tilde E}_0$.

For polar liquids, the function $\bm{\chi}_s(\mathbf{k})$ splits into
projections longitudinal (parallel) and transverse (perpendicular) to
the wave-vector $\mathbf{k}$. Each component is then represented by
the corresponding structure factor which is a function of the
magnitude of $k$ only.\cite{DMjcp2:04,DMcp:06} These structure factors
were obtained in this work from MD simulations of TIP3P
water\cite{tip3p:83} at different temperatures (see refs
\onlinecite{DMcp:06} and \onlinecite{DMjcp2:08} for more details).
With this input, the NRFT calculation is performed by grid summation
(in $\mathbf{k}$-space) of the solvent response function with the
solute electric field. This latter was calculated numerically by using
Fast Fourier Transform on the real-space lattice of $512^{3}$ points
with a grid spacing of $0.42$ \AA. In case of reorganization energy
calculations, $\mathbf{\tilde E}_0(\mathbf{k})$ in eq \ref{eq:10} is
obtained by taking only $\Delta z_j$ charges of the redox site, thus
producing the field $\Delta\mathbf{\tilde E}_0(\mathbf{k})$. In contrast,
the solvent component of the free energy gap of electron transfer, $\Delta
G_s$, is calculated with the complete charge distribution of the
protein.

\begin{figure}
  \centering
  \includegraphics*[width=6cm]{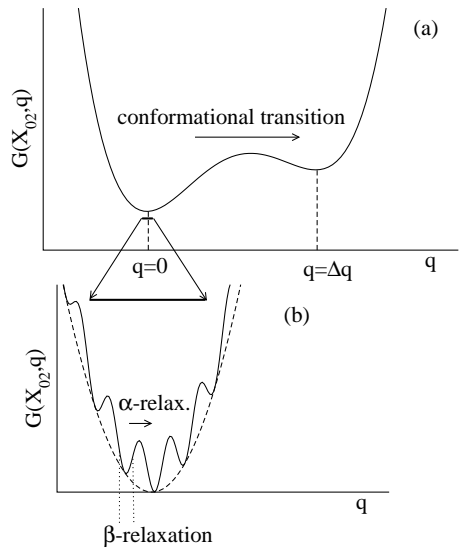}  
  \caption{Free energy landscape for conformational transitions (a)
    and $\alpha$-relaxation (b) of the protein. $G(X_{02},q)$ shows the
    cross-section of the two-dimensional free-energy surface taken at
    the equilibrium final value of the solvent reaction coordinate
    $X_{02}$. Each conformational state along the reaction coordinate
    $q$ contains a large number of substates separated by smaller free
    energy barriers (b). Transitions between these states are
    responsible for $\alpha$-relaxation of the protein slaved to water
    dynamics. Each of $\alpha$-substates can be separated into
    $\beta$-substates (not shown) responsible for $\beta$-relaxation of the
    protein and the hydration shell. }
  \label{fig:3}
\end{figure}

\section{Results}
\label{sec:3}
The conformational dynamics of proteins are very
disperse\cite{Frauenfelder:88} including several time-scales that can
potentially affect the energetics of electron transfer. Global
conformational changes of the protein, which often occur on the
time-scale of microseconds,\cite{Fenimore:02} make the slowest
time-scale. These transitions occur between minima of the free energy
landscape separated by highest barriers. In terms of the
two-dimensional coordinate space used in Figure \ref{fig:2}, this
motion sets up a transition from $q=0$ to $q=\Delta q$ along the
generalized protein coordinate $q$.  In Figure \ref{fig:3} we show the
cross-section of the free energy surface, $F(X_{02},q)$, at the final
state along the solvent polarization coordinate $X=X_{02}$.  The
activation barrier separating the states $q=0$ and $q=\Delta q$ is very
high to be observed on the time-scale of our MD simulations. The
topology of the free energy landscape\cite{Stillinger:95} is, however,
more complex than that is sketched in Figure \ref{fig:3}a. Each of the
conformational states, $q=0$ and $q=\Delta q$, contains a large number of
conformational substates separated by lower barriers\cite{Fenimore:04}
(Figure \ref{fig:3}b). Transitions between these substates represent
$\alpha$-relaxation of the protein with many features analogous to
$\alpha$-relaxation of structural glasses.\cite{Angell:95} These protein
dynamics are ``slaved'' to the solvent in a sense that the temperature
dependence of the corresponding relaxation time follows that of
water.\cite{Fenimore:04} One of the consequences of this slaving is
that the long-known dynamical transition of protein atomic
displacements above the linear regime at $T_{\mathrm{tr}}\simeq 200-250$
K\cite{Frauenfelder:88,Bizzarri:04,Angell:95} can be traced back to
the fragile-to-strong dynamic transition of hydrating
water.\cite{Caliskan:05,Chen:05} An alternative explanation suggests
the merger of the fast $\beta$- with slow non-observable $\alpha$-relaxation at
the transition temperature.\cite{Swenson:07} This fast $\beta$-relaxation
of the protein and hydrating water can be visualized as transitions
between low-barrier substates within each landscape basin of the
$\alpha$-relaxation processes (not shown in Figure \ref{fig:3}b).  Fast
fluctuations between these substates involve amino-acid side chains
and hydrogen-bond network at the protein
surface\cite{Tarek_PhysRevLett:02,Engler:03} as well as protein
vibrations which are not affected by the dynamic transition at
$T=T_{\mathrm{tr}}$. $\beta-$relaxation of the protein is strongly
dominated by $\beta$-relaxation of the hydrating water\cite{Fenimore:04}
involving translational motions of water molecules in and out of the
first hydration layer.\cite{Tarek_PhysRevLett:02}

\begin{figure}
  \centering
  \includegraphics*[width=6cm]{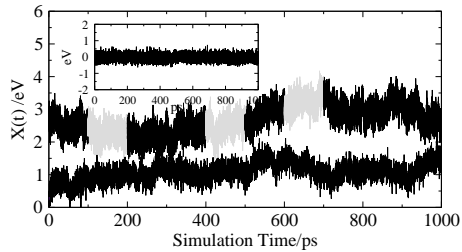}
  \caption{Trajectory of the solvent component of the donor-acceptor
    energy gap of PC in TIP3P water at 310 K. The upper trajectory
    shows unrestricted protein/water dynamics and the lower curve
    refers to protein atomic displacements frozen by applying positional
    harmonic restrains (8.0 kcal mol$^{-1}$ \AA$^{-2}$). Inset shows the
    same property for photoexcitation of tryptophan in TIP3P
    water. The gray regions in the upper trajectory indicate segments
    of the trajectory 100 ps long. The lower trajectory was shifted
    down by 1.5 eV for better visibility. }
  \label{fig:4}
\end{figure}

This scenario is consistent with the dynamics of the donor-acceptor
energy gap observed along the MD simulation trajectory. A
large-amplitude, redox-induced conformational transition, if it
exists,\cite{com:S07} is too slow to occur on the observation
timescale $\tau_{\mathrm{obs}}$ determined, in the computer experiment,
by the length of the simulation trajectory. However, both $\alpha$- and
$\beta$-relaxation of the protein and water (about 40\% of the overall
protein relaxation on the 10 ns time-scale\cite{McMahon:98}) are
clearly seen in the $X(t)$ trajectory (Figure \ref{fig:4}). The slower
$\alpha$-relaxation component is represented by large-amplitude
oscillations superimposed onto fast $\beta$-fluctuations of the solvent
dipoles. The slow $\alpha$-relaxation is not typically seen in small rigid
solutes exemplified by the trajectory of tryptophan (inset in Figure
\ref{fig:4}) where only fast $\beta$-fluctuations are present. The
time-scale of $\alpha$-fluctuations (ca.\ 1 ns) suggests their origin in
the motion of polar side groups\cite{Swenson:07} which show jumps in
their dihedral angles on the same time-scale.\cite{Baysal:05}

\subsection{Interactions with water and protein}
\label{sec:3-1}
The overall donor-acceptor energy gap $\Delta E$ is a sum of a gas-phase
component, mainly affected by the ligands coordinating copper, and
electrostatic interactions with the protein matrix, $\Delta
E_{\mathrm{prot}}$, and with the solvent (water), $X$. Their sum
makes the total reaction coordinate 
\begin{equation}
  \label{eq:11}
    Y= X + \Delta E_{\mathrm{prot}}  
\end{equation}
discussed in Sec.\ \ref{sec:2}.  In eq \ref{eq:11}, $\Delta
E_{\mathrm{prot}}$ is connected to characteristic conformational
coordinate of the protein $q$ by electron phonon coupling $\gamma$:
$E_{\mathrm{prot}}=\gamma q$. Table \ref{tab:1} lists the statistics of $Y$
fluctuations obtained from 10 ns of simulation data.  We report the
average donor-acceptor energy gap from the interaction of $\Delta z_j$
charges (Table S1) of the redox site (Red and Ox states) with water,
$\langle X\rangle$, and protein, $\langle\Delta E_{\mathrm{prot}}\rangle $, as well as their sum,
$\langle Y\rangle$. In addition, Table \ref{tab:1} gives the variance of $Y$ which
can be split into two self-correlation functions and one mixed
protein-water component:
\begin{equation}
  \label{eq:14}
  \sigma^2 = \langle(\delta Y)^2 \rangle  = \sigma_s^2 + \sigma_{\mathrm{prot}}^2 +
  \sigma_{s,\mathrm{prot}} 
\end{equation}

The splitting of the total variance into the reorganization energies
$\lambda_s = \sigma_s^2/(2k_{\text{B}}T)$ and $\lambda_{\mathrm{prot}} =
\sigma_{\mathrm{prot}}^2/(2k_{\text{B}}T)$ introduced in Sec.\ \ref{sec:0}
neglects the cross-correlation component $\sigma_{s,\mathrm{prot}}$ which, in our
simulations, amounts to 5--18\% of the total variance. This is the error
bar for the separation of protein and solvent nuclear fluctuations
into two separate stochastic processes.

\begin{table*}
  \caption{Averages (eV) and variances (eV$^2$)  
    of the interaction energy between the $\Delta z_j$ charges of the 
    active site and the solvent (s) and protein (prot) for PC 
    in Red and Ox states (eq  \ref{eq:14}). 
    $\langle Y\rangle $ and $\sigma^2$ refer to the total average energy gap and corresponding
    variance, respectively. The data are collected from a $10$ ns trajectory.  }
\label{tab:1}
\begin{tabular}{ccccccccc}
\hline
 $T/K$ & Redox state & $\langle X \rangle $ & $\langle\Delta E_{\mathrm{prot}}\rangle$ & $\langle Y\rangle$ & 
   $\sigma_{\mathrm{prot}}^{2}$ & $\sigma_{s}^{2}$ & $\sigma^{2}$ & $\sigma_{\mathrm{prot},s}$ \\ 
\hline 
$310$ & Ox & $2.484$ & $-8.587$ & $-6.103$ & $0.085$ & $0.290$ & $0.275$ & $-0.051$ \\
      & Red & $3.394$ & $-8.690$ & $-5.296$ & $0.078$ & $0.298$ & $0.416$ & $0.014$ \\
$285$ & Ox & $2.460$ & $-8.465$ & $-6.006$ & $0.082$ & $0.268$ & $0.287$ & $-0.053$ \\
      & Red & $3.773$ & $-8.964$ & $-5.190$ & $0.071$ & $0.233$ & $0.249$ & $-0.028$ \\
\end{tabular}
\end{table*}

\subsection{Solvent reorganization energy}
\label{sec:4}
Reorganization energy $\lambda_q(\tau_{\mathrm{obs}})$ in eq \ref{eq:3}
originates from fluctuations of the solvent dipolar polarization
induced by coupled protein-solvent dynamics.  This reorganization
energy thus represents a new solvent mode affecting the donor-acceptor
energy gap absent in the traditional theories of electron transfer
operating in terms of separate vibrational ($\lambda_i$) and polarization
($\lambda_p$) nuclear modes. The enhancement of the energy gap variance by
this new mode is very significant: the variance of $X$ changes 
from $\sigma_s^2=2k_{\mathrm{B}}T\lambda_p$ ($\lambda_p=\Delta X/2 \simeq 0.45-0.65$ eV),
comparable to other simulations,\cite{Cascella:06,Simonson:02} to a
much higher value $\sigma_s^2=2k_{\mathrm{B}}T\lambda_s$ characterized by the
solvent reorganization energy $\lambda_s = \lambda_p + \lambda_q \simeq 5$ eV (Table
\ref{tab:2}).  This gigantic value of the reorganization energy far
exceeds what is typically observed for electron transfer reactions
between small hydrated ions.\cite{Kuharski:88} 

The variance of the donor-acceptor energy gap $\sigma_q \simeq 0.5$ eV produced
by conformational flexibility in our simulations is significantly
higher than experimentally reported $\sigma_q \simeq 0.05$ eV from charge
recombination in bacterial reaction centers trapped by cooling (eq
\ref{eq:7})in their conformational substates.\cite{McMahon:98} As
noted above, this lower variance of energy gaps in reaction centers is
expected since water dynamics significantly contributing to
fluctuations of the donor-acceptor energy gap is also quenched by
cooling. In addition, the hydrophobic environment of cofactors located
in the membrane protein complex and the low temperature of the kinetic
arrest $T_{\mathrm{kin}}\simeq 175$ K (eq \ref{eq:7}) both contribute to
the lower $\sigma_q$.

\begin{table}
\caption{Reorganization parameters of PC (all energies are in eV). }
\label{tab:2}
\begin{tabular}{lllll}
\hline
$T$(K) & $\lambda_p$ & $\lambda_q$ & $\Delta G_s$\footnotemark[1] & $\lambda_{\mathrm{prot}}$\footnotemark[2] \\
\hline
285    & 0.81\footnotemark[3]            &  4.8(Red)   &  $-3.2$       & 1.6  \\
       & 0.77(0.69,3.6)\footnotemark[4]  &  4.1(Ox)    &  $-4.7(-7.1,-9.6)$\footnotemark[4]    &         \\
310    & 0.54\footnotemark[3]            &  5.0(Red)   & $-2.9$ & 1.5  \\
       & 0.74(0.54,3.6)\footnotemark[4]  &  4.5(Ox)    & $-4.6(-7.1,-9.6)$\footnotemark[4] &           \\
\end{tabular}
\footnotetext[1]{Solvent component of the redox free energy obtained from the simulation
  data as $\Delta G_s = G_s^{\mathrm{Red}} - G_s^{\mathrm{Ox}}=-(\langle X\rangle_1 + \langle X\rangle_2)/2$.}
\footnotetext[2]{Calculated from the variance $\sigma_{\mathrm{prot}}^2$ of the electrostatic interaction
  of $\Delta z_j$ charges of the redox site with the charges of the protein as
  $\lambda_{\mathrm{prot}} = \sigma_{\mathrm{prot}}^2/(2k_{\mathrm{B}}T)$.}
\footnotetext[3]{Calculated from the simulation data as $(\langle X\rangle_1-\langle X\rangle_2)/2$.} 
\footnotetext[4]{Theoretical calculations using atomic charges, vdW radii, and 
  coordinates of the protein combined with microscopic, non-local response functions of 
  water.\cite{DMjcp2:04,DMcp:06} Dielectric continuum calculations
  (brackets) have been done with the solvent-accessible cavity (first
  number) and standard vdW cavity (second number).}
\end{table}

We need to emphasize that the reorganization energies considered
here refer to the change in the charge distribution of PC only.  A
donor-acceptor complex composed of two proteins (photosynthetic
electron transfer) will also include a change of charges on the
partner (heme) protein. The interaction of this other set of charges
located within a membrane protein with water is expected to be weaker
than for hydrated PC.  Therefore, there should be only minor Coulomb
correction to the reorganization energy.  In addition, some reduction
of the reorganization energy will arise from electronic polarizability
of water not included in TIP3P parametrization (our calculations using
non-local solvent response\cite{DMjcp2:04,DMcp:06} show a reduction
from 0.74 eV in TIP3P water to 0.40 eV in ambient water).
Nevertheless, the gigantic magnitude of $\lambda_p+\lambda_q$ compared to the
commonly considered $\lambda_p$ calls for attention to the effects of
coupled protein/water dynamics on electron transfer.

\begin{figure}
  \centering
  \includegraphics*[width=6cm]{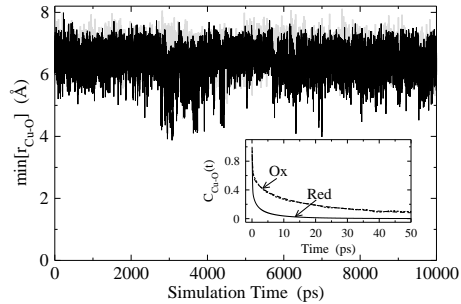}
  \caption{Trajectory of the closest distance between Cu of the
    plastocyanin active site (Figure \ref{fig:1}) and oxygen of water
    in the Red (gray) and Ox (black) states. The inset shows time
    autocorrelation functions of the Cu-O distance in Red and Ox
    states. The autocorrelation functions fit well to eq
    \ref{eq:5} with the set of fitting parameters $\{A_G,\tau_G,\tau_E,\beta\}$:
    $\{0.31,0.2,12.3,0.53\}$ for Ox and $\{0.39,0.2,1.1,0.46\}$ for Red.
    Here, $\tau_G$ and $\tau_E$ are in picoseconds. }
\label{fig:5}
\end{figure}

The solvent effect on the electron transfer thermodynamics is
dominated by water molecules closest to the active site.  Protein
flexibility significantly modulates this first solvation shell
producing fluctuations of the closest Cu-O distance around the average
of 6.64 \AA, the largest fluctuation amplitude of $\simeq 2$ \AA, and the
standard deviation of 0.6 \AA{} (Figure \ref{fig:5}).  We also note that
the Cu-O pair distribution function (Figure S2) does not change with
changing the redox state of plastocyanin, in contrast to observations
reported for heme
proteins.\cite{Cascella:06,Blumberger:06,Sulpizi:07,Simonson:02} With
such large-amplitude fluctuations, water is effectively further apart
from the protein surface than the distance of the closest approach.
The solvent part of the reaction free energy is then nearly 1.5 times
smaller in magnitude than the value calculated from our solvation
model\cite{DMjcp2:04,DMcp:06} assuming the closest approach (Table
\ref{tab:2}).  In contrast, the calculated reorganization energy $\lambda_p$
is in good agreement with simulations, which supports our assumption,
used to derive eq \ref{eq:3}, that conformational fluctuations do not
affect this parameter.

We note in passing that $\lambda_p$ from continuum calculations is very
sensitive to the definition of the dielectric cavity. When van der
Waals (vdW) radii of protein atoms are used to determine the cavity,
high-polarity dielectric is allowed in a narrow pocket near copper
thus significantly increasing the free energy of solvation. When, in
contrast, a water molecule is rolled on the vdW surface to determine
the solvent-accessible cavity, the result of continuum calculations is
comparable to both the NRFT and MD numbers. We will provide a more
detailed discussion of these results in a separate
publication.\cite{DMjcp2:08}

The large value of $\lambda_q$ raises the question of whether the new
nuclear mode responsible for the energy gap variation should be
attributed solely to conformational motions of the protein or to more
complex collective dynamics coupling solvent to protein
fluctuations. The evidence existing in the literature advocates the
latter view suggesting that both the $\alpha$- and $\beta$-relaxation of the
protein are strongly coupled to hydrating water. Our attempts to
connect the slow modulations of the $X(t)$ trajectory (Figure
\ref{fig:4}) to the vibrational density of states of the
protein\cite{Go:83} have not given positive results since the
low-frequency vibrations seen in Figure \ref{fig:4} could not be
resolved from the quasi-harmonic analysis\cite{Go:83} (Figure S1) or
from the intermediate scattering function.\cite{Bizzarri:04} We have
also tried to freeze the protein motions through harmonic positional
restraints on atomic translations, with the restraint weight equal to
8.0 kcal mol$^{-1}$ \AA$^{-2}$. These simulations (ca.\ 5 ns started at
the end of the unrestrained trajectory) have resulted in the energy-gap
variance $\sigma^2$ diminished by a factor of $\simeq 3$ (lower trajectory
in Figure \ref{fig:4}), but still not reaching the value $\sigma_p^2$ from
the Stokes shift. This observation supports the view that
$\alpha$-fluctuations of the donor-acceptor gap are coupled to
translational motions within the hydration layer at the protein
surface and that these fluctuations cannot be separated from protein's
conformational dynamics. Nevertheless, the strong reduction of
$\sigma^2$ upon freezing of the protein still suggests that protein
motions produce the largest energetic contribution to the
reorganization energy $\lambda_q$.

\subsection{Protein dynamics}
\label{sec:4-1}
\begin{figure}
  \centering
  \includegraphics*[width=6cm]{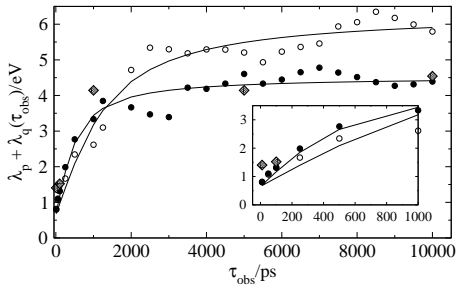}
  \caption{Solvent (water) reorganization energy of Ox (open circles)
    and Red (filled circles) states of PC at 285 K vs the observation
    time $\tau_{\mathrm{obs}}$ defined as the length of trajectory over
    which the averages were calculated.  The solid lines are fits of
    simulations to eq (\ref{eq:4}) with $\tau_q=1$ ns (Ox) and
    $\tau_q=0.5$ ns (Red). The hatched diamonds indicate the results from
    ref \onlinecite{Blumberger:06}. The inset shows the initial portion of
    the plot. }
\label{fig:6}
\end{figure}

If the large-amplitude protein/water motions affecting the solvent
polarization are overdamped,\cite{Markelz:07} they can be described by
Debye relaxation with an effective relaxation time $\tau_q$.  The use of
the Debye relaxation function in eq \ref{eq:8} gives the following
simple equation for the non-ergodic reorganization
energy:\cite{DMjcp2:06}
\begin{equation}
  \label{eq:4}
   \lambda_q(\tau_{\mathrm{obs}}) =(2\lambda_q/ \pi) \mathrm{cot}^{-1}(\tau_q/\tau_{\mathrm{obs}} ) 
\end{equation}
Protein dynamics coupled with dipolar solvent polarization are
dominated by very slow motions with the characteristic time $\tau_q$ of
about 0.5--1 ns, as follows from the fit of $\lambda_p +
\lambda_q(\tau_{\mathrm{obs}})$ (eq \ref{eq:4}) to the simulation data (Figure
\ref{fig:6}).  The non-ergodic component $\lambda_q(\tau_{\mathrm{obs}})$ was
calculated from the energy gap variance taken on observation windows
$\tau_{\mathrm{obs}}$ along the simulation trajectory (gray segments of
length 100 ps in Figure \ref{fig:4}). On short observation times,
$\tau_{\mathrm{obs}}< 100$ ps, i.e.  fast electron transfer reactions,
the slow conformational modulation does not show up, and the
reorganization energy from the width approaches that from the Stokes
shift, thus restoring eq \ref{eq:1}. A similar behavior, including the
magnitude of the corresponding reorganization energy, was observed in
ref \onlinecite{Blumberger:06} (hatched diamonds in Figure
\ref{fig:6}).

With such slow conformational modulation of the water polarization,
each short segment of the long trajectory finds itself in a different
configuration, a situation akin to dynamical heterogeneity responsible
for stretched-exponential relaxation of structural
glasses.\cite{Angell:95} This picture is indeed confirmed by the
Stokes shift correlation function calculated on segments of trajectory
of different length.  Stokes shift correlation function $C(t)=\langle X(t)
X(0)\rangle$ from a short segment has a typical biphasic form composed of a
fast Gaussian decay followed by exponential
relaxation\cite{Jimenez:94} for which the stretch exponent $\beta$ in eq
\ref{eq:5} is equal to unity:
\begin{equation}
  \label{eq:5}
  C(t) = A_G e^{-(t/\tau_G)^2} + (1 - A_G)e^{-(t/\tau_E)^{\beta}} 
\end{equation}
On the contrary, the Stokes shift correlation function calculated on a
longer segment (1--2 ns) develops a stretched-exponential relaxation with
the stretching exponent $\beta=0.69$ and relaxation time of about 150 ps
(Figure S3).  This long tail, which may require longer simulations to
be fully resolved,\cite{McMahon:98} is caused by collective water
displacements (Figure \ref{fig:1}) by slowly moving parts of a
biopolymer.\cite{Nilsson:05,Berg:05}

\subsection{Free energy surfaces}
\begin{figure}[t]
  \centering
  \includegraphics*[width=6cm]{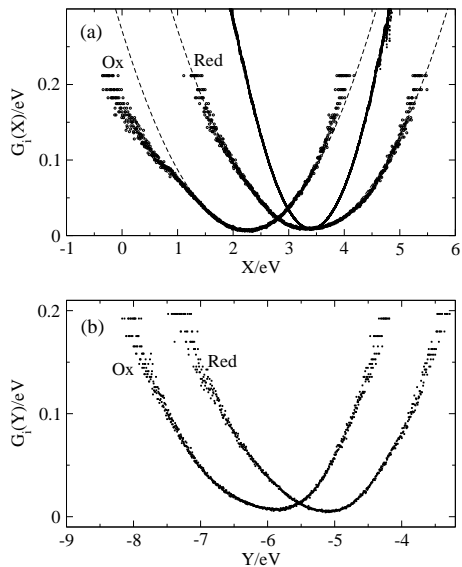}
  \caption{Electron transfer free energy surfaces of PC at 310 K
    plotted against the solvent reaction coordinate (a) and against
    the total (water+protein) reaction coordinate $Y$ (b).  The dashed
    lines in A are fits of $G_i(X)$ from the 10 ns trajectory to eq
    (\ref{eq:3}). The narrow curve in (a) has been obtained by
    calculating the distribution functions on 100 ps segments of the
    trajectory (gray segments in Figure \ref{fig:4}) and averaging them
    after sliding to a common maximum.  All curves are logarithms of
    normalized distribution functions along the corresponding reaction
    coordinate.  }
  \label{fig:7}
\end{figure}

The picture of non-ergodic, glassy dynamics emerging from the static
and time-resolved energy-gap statistics is consistent with the free
energy surfaces $G_i(X)$. They are very shallow on the long 10 ns
trajectory becoming increasingly curved on a shorter observation
window (Figure \ref{fig:7}a). The full free energy surfaces, obtained
by sampling the total interaction energy of the active site with both
the protein and the solvent (reaction coordinate $Y$ in Figure
\ref{fig:2}), are uniformly shifted to the negative values of $Y$
without significant change in the Stokes shift (Table \ref{tab:1} and
Figure \ref{fig:7}b).  The slow non-ergodic dynamics of protein
conformations thus decouples the Stokes shift from the reorganization
energy breaking eq \ref{eq:1} down. Accordingly, the difference
$G_2(X) - G_1(X)$ in the region of overlap of Ox and Red surfaces is
still a linear function of $X$, but with the slope of 0.10, instead of
the unitary slope predicted by Marcus theory and usually observed in
fully equilibrated systems.\cite{Blumberger:06} This number is
consistent with eq \ref{eq:3} which yields the slope of $\Delta X/(\Delta
X+2\lambda_q)\simeq 0.13$.

The statistics of the donor-acceptor energy gap induced by protein
fluctuations are also approximately Gaussian.  The width of the
distribution is given by the variance of the electrostatic interaction
energy of $\Delta z_j$ charges of the redox site with the protein atomic
charges.  The corresponding reorganization energy is approximately 1.5
eV (Table \ref{tab:2}). However, since the reaction path is expected
to follow the fast solvent coordinate, our main focus here is on the
free energies $G_i(X)$.

These results do not contradict experimental estimates of $\lambda_p+\lambda_i \simeq
0.6-0.8$ eV for copper proteins obtained from the top of the energy
gap law when the reaction barrier disappears.\cite{Gray:00,Skov:98}
From eq \ref{eq:3}, the activationless transition is achieved at
$X_{0i}=0$ when the reaction free energy $\Delta G_0$ obeys the equation
$-\Delta G_0 - \lambda_i = \Delta X/2=\lambda_p$.  With the inner reorganization energy
currently estimated as low as 0.1 eV,\cite{Cascella:06}
$\lambda_p=0.54-0.81$ eV from present simulations is consistent with
experiment. On the other hand, reorganization energy $\lambda_s$ entering
the distribution width $\sigma_s^2$ is about an order of magnitude higher.
The breakdown of the link between the Stokes shift and the
distribution width (eq \ref{eq:1}) must have significant implications
for the biological function of PC and probably of other electron carrier
proteins. The large distribution width leads to an extremely small
activation barrier, $\Delta G^{\mathrm{act}} = (\lambda_p + \Delta G_0)^2/(4\lambda_p +
4\lambda_q) \simeq 0.08$ eV, and thus fast electron transfer, for reactions at
the docking locations with the typically small reaction free energy $\Delta
G_0 \simeq -20$ meV.\cite{Ubbink:98} The standard rate estimate
\cite{Noy:06} then gives 13 \AA{} for the donor-acceptor distance at which
the threshold catalytic rate of $10^4$ s$^{-1}$ is achieved.

\section{Concluding Remarks}
The application of the ideas presented here to biological electron
transfer requires the transition from the observation time determined by
the length of the simulation trajectory to the reaction kinetics. This
is achieved by setting $\tau_{\mathrm{obs}}=\tau_{\mathrm{ET}}$ which, given
the activation barrier is a function of $\tau_{\mathrm{obs}}$, leads to a
self-consistent equation for the rate \cite{DMacc:07}
\begin{equation}
  \label{eq:6}
     k_{\mathrm{ET}} \propto \exp\left[ - \Delta G^{\mathrm{act}}(k_{\mathrm{ET}})/ (k_{\mathrm{B}}T) \right] 
\end{equation}

\begin{figure}
  \centering
  \includegraphics*[width=6cm]{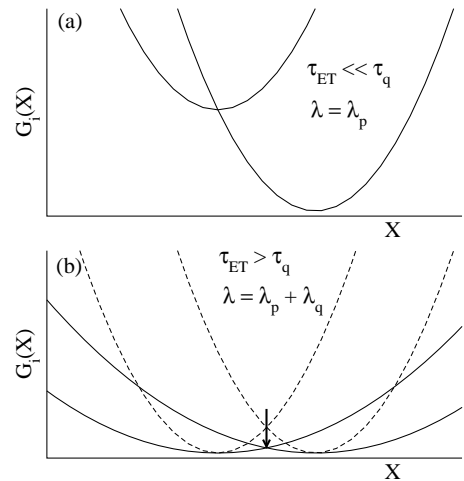}
  \caption{Energetics of fast electron transfer reactions losing in
    redox potential to achieve activationless transitions (a) and of
    slower (ns range) reactions (b) which are allowed to proceed with
    a small driving force. The reaction barrier is lowered in the
    latter case by non-ergodic conformational/water dynamics
    transforming the dashed-line parabolas into solid-line parabolas
    (b).  The vertical arrow in (b) shows the suppression of the
    activation barrier by reorganization energy $\lambda_q$. }
  \label{fig:8}
\end{figure}

Equation \ref{eq:6}, incorporating the notion of reaction
non-ergodicity, offers a compelling picture of the hierarchy of
electron transfer reactions in photosynthetic systems.  Faster
reactions with $k_{\mathrm{ET}}\gg 10^9$ s effectively cut off slow
conformational motions of the protein from their energetics (Figure
\ref{fig:6}), and the standard picture \cite{MarcusSutin} of the
Stokes shift and parabolas' curvature related by eq \ref{eq:1}
applies. Realizing fast electron transfer then requires activationless
transition as observed in primary charge separation in photosynthetic
reaction centers (Figure \ref{fig:8}a).\cite{BixonJortner:99} On the
contrary, reactions in the sub-nanosecond range start to experience
the effect of conformational modulation of the activation barrier and
can in fact proceed efficiently even with a small driving force since
the activation barrier is lowered by the growing width of the energy
gap distribution (Figure \ref{fig:8}b).  Therefore, when speed is at
stake, natural systems have to lose in redox potential in exchange for
fast activationless transitions.  When slower reactions, still faster
than catalytic rates, can be afforded, losing reduction potential is
not the necessity, and reactions with low driving force can still be
efficient. 

Strong coupling between dipolar polarization and protein mobility
advocated here is consistent with the long suggested connection
between protein dynamics and
hydration.\cite{Tarek_PhysRevLett:02,Fenimore:04,Caliskan:05,Chen:05,Swenson:07}
The new reorganization energy discovered here is related to the rms
displacement of the slow mode as $\lambda_q \propto \langle(\delta q)^2\rangle/T$ and is therefore
expected to show a sharp increase at $T>T_{\mathrm{tr}}$ when $\langle(\delta
q)^2\rangle$ starts its nonlinear rise. This observation might provide a
resolution of the long-standing puzzle of electron transfer kinetics
in many plants and bacteria: Arrhenius plots of electron transfer
rates often show breaks in their slopes at temperatures consistent
with $T_{\mathrm{tr}}$. \cite{Vault:66,Hales:76,Borovikh:05} This
feature might be linked to the rise of $\lambda_q$ at the onset of
conformational activity in proteins.\cite{Parak:03}

The existence of the solvent-slaved $\alpha$-relaxation is presently traced
back to strong coupling between polar/ionized surface residues and
water. This type of dynamics is usually not observed in inorganic and
organic donor acceptor complexes used for ET reactions and is
presently believed to be unique to natural polymers. Properties of
synthetic polymers, in particular in respect to glassy dynamics,\cite{Frick:95} have
many common features with biopolymers and one hopes that phenomena
analogous to ones observed here might be realized in flexible
donor-acceptor architectures including branched polymers and
dendrimeric structures.

\begin{acknowledgments}
  This work was supported by the NSF (CHE-0616646). CPU time was
  provided by Pittsburgh Supercomputer Center and ASU's Center for
  High Performance Computing. We are grateful to Marshall Newton for
  critical comments on the manuscript.
\end{acknowledgments}

\bibliographystyle{achemso}

\bibliography{/home/dmitry/p/bib/chem_abbr,/home/dmitry/p/bib/photosynthNew,/home/dmitry/p/bib/glass,/home/dmitry/p/bib/et,/home/dmitry/p/bib/dm,/home/dmitry/p/bib/protein,/home/dmitry/p/bib/solvation,/home/dmitry/p/bib/bioet,/home/dmitry/p/bib/etnonlin,/home/dmitry/p/bib/dynamics,/home/dmitry/p/bib/photosynth,/home/dmitry/p/bib/simulations,/home/dmitry/p/bib/liquids}

\end{document}